\documentclass[fleqn,usenatbib]{mnras}

\usepackage{newtxtext,newtxmath}

\usepackage[T1]{fontenc}

\DeclareRobustCommand{\VAN}[3]{#2}
\let\VANthebibliography\thebibliography
\def\thebibliography{\DeclareRobustCommand{\VAN}[3]{##3}\VANthebibliography}

\usepackage{graphicx}	
\usepackage{amsmath}	
\usepackage{verbatim}
\usepackage[detect-all]{siunitx}
\usepackage{comment}

\title[Irradiation effects on self-gravitating discs]{Effect of irradiation model on 2D hydrodynamic simulations of self-gravitating protoplanetary discs}

\author[C. S. Leedham et al.]{
Caitriona S. Leedham,$^{1}$\thanks{E-mail: \url{cscl3@cam.ac.uk}}
Richard A. Booth,$^{2}$
Cathie. J. Clarke$^{1}$
\\
$^{1}$Institute of Astronomy, University of Cambridge, Madingley Road, Cambridge, CB3 0HA, UK\\
$^{2}$School of Physics and Astronomy, University of Leeds, Leeds, LS2 9JT, UK
}

\date{Accepted XXX. Received YYY; in original form ZZZ}

\pubyear{2024}

\begin{document}
\label{firstpage}
\pagerange{\pageref{firstpage}--\pageref{lastpage}}
\maketitle

\begin{abstract}
Young protoplanetary discs are expected to be gravitationally unstable, which can drive angular momentum transport as well as be a potential mechanism for planet formation. Gravitational instability is most prevalent in the outer disc where cooling timescales are short. At large radii, stellar irradiation makes a significant contribution to disc heating and is expected to suppress instability. In this study, we compare two models of implementing irradiation in 2D hydrodynamic simulations of self-gravitating discs: supplying a constant heating rate per unit mass and per unit area of the disc. In the former case, instability is quenched once the stellar irradiation becomes the dominant heating source. In the latter case, we find instability persists under high levels of irradiation, despite large values of the Toomre Q parameter, in agreement with analytic predictions. Fragmentation was able to occur in this regime with the critical cooling timescale required decreasing as irradiation is increased, corresponding to a maximum threshold for the viscosity parameter: $\alpha\sim0.03-0.09$.
\end{abstract}




\section{Introduction}
\label{sec:intro} 

Protoplanetary discs form around young stars as infalling material is flattened into a disc due to angular momentum conservation. Young discs are dominated by their gas component and can be massive enough to be self-gravitating. The destabilising force of gravity can overcome support from pressure and centrifugal forces, causing regions of gas to clump together. The potential for gravitational instability (GI) has considerable implications for the evolution of the disc and planet formation. The formation of bound objects via the runaway collapse of overdensities (fragmentation) has been proposed to explain the high occurrence rate of binary stars \citep{Clarke2009} and to account for the formation of giant planets at large orbital radii \citep{Boss1997}. Additionally, non-axisymmetric structures induced by GI, such as spiral arms \citep{Lynden-Bell1972}, facilitate angular momentum transport -- a critical process in accretion disc theory that enables material to flow inward.

Previous studies have emphasised that an important role for GI, both in terms of planet formation and driving significant angular momentum transfer, relates to the ratio of the cooling time to the dynamical time. Consequently, the regions where GI is potentially important are limited to the outer parts of the disc ($r\gtrsim40AU$, \cite{Rafikov2005,Matzner2005,Stamatellos2008,Clarke2009,Forgan2011}). However, these regions are precisely where it can be expected that external heating of the disc by stellar irradiation is significant \citep{D'Alessio1999,Kratter_Lodato2016}. The regime of short cooling time and strong irradiation is therefore an important one to understand but previous studies are unclear about the behaviour of the disc in this region. In this paper we focus on simulations that explore this regime in detail.

The linear theory described by \cite{Toomre1964} applies perturbations to a 2D self-gravitating axisymmetric rotating disc. The onset of instability is characterised by the Toomre parameter:

\begin{equation} \label{eq:Toomre}
    Q=\frac{c_s\Omega}{\pi G\Sigma},
\end{equation}

\noindent which represents the balance of self-gravity against pressure and centrifugal forces. Low and less stable values of $Q$ are found at large radii where the disc is cooler. Classically, $Q<1$ is required for instability, however, realistic discs will undergo heating and cooling processes that allow them to thermally saturate at higher values of $Q$. As regions of the disc collapse, $Q$ decreases and the increased heating rate due to shock dissipation will stabilise overdensities. As they expand, the heating rate decreases and the disc begins to cool again. This self-regulated state \citep{Paczynski1978} is known as gravitoturbulence and numerical simulations show discs saturate with $Q\approx 1-2$ \citep{Lodato2004,Rice2011}.

Gravitoturbulence is sensitive to the thermodynamics of the disc. Heating due to shocks acts to increase $Q$, whereas radiative cooling will decrease $Q$ towards instability. The more efficient the cooling, the higher the rate of heating required to balance it. Cooling is often implemented using the $\beta$--prescription \citep{Gammie2001}, where $\beta=\tau_{\mathrm{c}}\Omega$ is the ratio of the cooling timescale of the disc material to the orbital timescale. $\beta$ is a decreasing function of radius such that larger radii have shorter cooling times \citep{Rafikov2005,Clarke2009}. The effective viscosity can be modelled by the \cite{Shakura_Sunyaev1973} prescription, $\nu = \alpha c_s H$, where $\alpha$ is a dimensionless parameter measuring the efficiency of angular momentum transport due to turbulence. Assuming thermal equilibrium in the saturated state leads to an inverse relationship between $\alpha$ and $\beta$: $\alpha=\frac{4}{9 \gamma(\gamma-1) \beta}$, derived by \cite{Gammie2001} and similarly by \cite{Mamatsashvili2009} using the total energy equation. This relation assumes the energy associated with angular momentum transport is dissipated locally, which has been tested numerically using global simulations and shown to be a good approximation \citep{Lodato2004,Lodato2005,Cossins2009}.

A further source of heating is irradiation from the central star. While incident flux decreases with orbital radius, it does so less steeply than internal heating in the disc \citep{Kratter_Lodato2016}. At large radii, it becomes a significant contribution to the overall heating. High levels of irradiation are expected to stabilise the disc by raising the value of $Q$. However, the outer disc is also where the cooling timescale becomes short, making the disc more susceptible to fragmentation. Thus it is import to understand the role of irradiation in these outer regions and whether they do, or do not, satisfy conditions required for planet formation by GI.

How irradiation is incorporated in realistic disc simulations is complicated by optical depth effects and the computational cost of radiative transfer models. A common simplification used in 2D local disc models is to include a constant added rate of energy determined by the stellar luminosity and radial location in the disc \citep{Rice2011,Baehr2015} to the energy equation. Irradiation creates an effective heat bath and imposes a background sound speed. It is expected that once this sound speed implies $Q>1$, a gravitoturbulent state cannot be maintained.

This additional term has been treated in various ways in previous studies with a clear difference being whether or not the heating rate per unit area depends on surface density. Linear analysis studies show distinctly different behaviours depending on the choice made \citep{Lin_Kratter2016}. Density dependence provides an extra stabilising term such that at large levels of irradiation implying $Q\gtrsim1-2$, there are no unstable solutions. Alternatively, without it, unstable solutions remain for arbitrarily high irradiation. This has been seen in numerical simulations, such as \cite{Rice2011}, who simulate the first case and find no evidence of instability for highly irradiated discs. However, using the latter case, \cite{Lohnert2020} find active GI for high levels of irradiation, limited only by numerical effects.

A further significant outcome of this case is the value of Q obtained in the gravitoturbulent state. \cite{Lohnert2020} found that as irradiation is increased, the saturated value of Q also increases and GI still occurred for discs with $Q\sim 10$. This is considerably different from the classical expectation that only $Q$ close to 1 can be unstable. Measurements of disc masses show that there are many discs not satisfying $Q<1$, but would instead meet this more relaxed criterion \citep{Tobin2020}.

Since the energy equation describes the evolution of the internal energy per unit area, including irradiation as a constant additional term corresponds to a constant heating rate of the disc per unit area. If the added term has a linear dependence on surface density, this is equivalent to a constant heating rate per unit mass with denser regions being supplied more energy per time. The outcome of this difference can be understood by considering that GI involves the continuous formation and disruption of overdensities. This requires self-gravity to dominate over stabilising effects, such as irradiation and enhance the growth of density fluctuations. When a constant rate of energy is supplied per unit mass, overdense regions contain more mass per area and will preferentially heated. This additional pressure support acts against self-gravity in promoting the growth of overdensities, but does not occur when heating per unit area.

In the outer disc where GI is relevant, opacity is dominated by icy grains and the disc is expected to be optically thick to incident stellar irradiation \citep{Clarke2009}. Therefore, it is more physically reasonable to incorporate heating as a constant rate over a surface area. Simulations in the literature more commonly model the case of constant heating per unit mass \citep{Rice2011,Baehr2015,Baehr2017} and do not explore the strongly irradiated regime (where Q > 1) since these are expected to be stable. \cite{Lohnert2020} simulate this regime using a heating per area model, but only consider the implications on discs with long cooling timescales, which would not be susceptible to fragmentation ($\beta\gg3$).

In this paper, we study both methods of heating due to irradiation in local 2D hydrodynamic simulations and compare the results to the expectations of the linear theory. We also investigate the boundary for fragmentation as a function of the cooling time and irradiation. Sections \ref{sec:Analytic Considerations} and \ref{sec:Methods} describe the analytic theory and simulations respectively. We present the results in Section \ref{sec:results} and our conclusions in Section \ref{sec:Conclusion}.

\section{Analytic Considerations}
\label{sec:Analytic Considerations}

In this section we set out the linear theory of self-gravitating discs with simply prescribed cooling and irradiation and discuss the requirements imposed by thermal equilibrium on the properties of the saturated state. The theory presented here will be useful in interpreting the simulations presented in Section \ref{sec:results}, which compare the outcome of different irradiation models. It is useful first to introduce terminology that will be used throughout the remainder of the paper.

$Q_{\text{irr}}=\frac{ c_{s,\text{irr}}\Omega_0}{\pi G \Sigma_0}$ is the value of the Toomre Q parameter that the system would have if it were subject to external irradiation with no contribution from heating by GI. $c_{s,\text{irr}}$ is the sound speed set by irradiation and $\Sigma_0$ is the background density. $Q_{\text{irr}}$ is thus an independent parameter that in Section \ref{sec:Methods} is varied between simulations.

$Q_{\text{sat}}=\langle Q \rangle$ is the value of the Toomre Q parameter that the system attains in practice when subject to external irradiation, cooling and heating effects associated with GI. $\langle \rangle$ denotes an average over the simulation domain and time.

$Q_{\text{sat,0}}=Q_{\text{sat}}(Q_{\text{irr}}=0)$ is the value of the Toomre Q parameter that the system attains in practice when subject to cooling and heating effects associated with GI but in the absence of external irradiation.

\subsection{The Model}
\label{sec:Model} 

We use the local, two-dimensional shearing sheet approximation \citep{Goldreich1965,Hawley1995} to model the disc with coordinates $(x,y)$ representing the radial and azimuthal directions respectively. The equations of hydrodynamics are solved in a frame co-rotating with the background Keplerian flow. They include the  mass continuity,  momentum conservation and internal energy density equations and a Poisson equation for the self-gravity of a razor-thin disc.

\begin{equation} \label{eq:continuity_eqn}
\partial_t \Sigma+\nabla \cdot(\Sigma \mathbf{u})=0
\end{equation}

\begin{equation} \label{eq:mtm_eqn}
\partial_t \mathbf{u}+(\mathbf{u} \cdot \nabla )\mathbf{u}=-\frac{1}{\Sigma} \boldsymbol{\nabla} P -2 \Omega_0\mathbf{e}_z \times \mathbf{u} + 3\Omega^2_0 x \mathbf{e}_{x} -\nabla \Phi    
\end{equation}

\begin{equation} \label{eq:Energy_eq}
\partial_t U+(\mathbf{u} \cdot \nabla) U + \gamma U(\nabla \cdot \mathbf{u})=-\frac{U}{\tau_{\mathrm{c}}} + \frac{U_\text{irr}}{\tau_{\mathrm{c}}}\left(\frac{\Sigma}{\Sigma_0}\right)^{f_\theta}
\end{equation}

\begin{equation} \label{eq:Poisson_eqn}
\nabla^{2} \Phi=4 \pi G \Sigma \delta(z)
\end{equation}

The forces on the right-hand side of equation \eqref{eq:mtm_eqn} are the pressure gradient, the Coriolis and tidal forces due to the rotating frame and the self-gravitational force. $\mathbf{u}$ is the velocity of the fluid relative to the background flow ($\mathbf{u}_0 = -\frac{3}{2}\Omega_0 x \mathbf{e}_{y}$). In the 2D approximation, $\Sigma = \int \rho dz$ is the vertically integrated density and similarly $P$ is the vertically integrated pressure. 

$U$ is the internal energy density (energy per unit area) and is related to the pressure by the equation of state, $P=(\gamma -1)U$. $U$ evolves via equation \eqref{eq:Energy_eq}, where the first term on the right-hand side represents radiative cooling with cooling time $\tau_{\mathrm{c}}=\beta / \Omega_0$. The second term is heating due to irradiation. $U_{\text{irr}}=\frac{\Sigma_0c_{\text{s,irr}}^2}{\gamma(\gamma-1)}$ indicates the internal energy per unit area for an isothermal disc that is heated to a sound speed $c_s=c_{\text{s,irr}}$ by an external irradiation field (in the absence of self-gravity), where $c_s^2=\gamma\frac{P}{\Sigma}$ is the adiabatic sound speed. Thus, $U_{\text{irr}}$ or $c_{\text{s,irr}}$ parameterises the strength of external heating in terms of this equilibrium temperature. 

It can be seen that the heating prescription in equation \eqref{eq:Energy_eq} corresponds to constant heating per area for $f_\theta=0$ and constant heating per mass for $f_\theta=1$. Note these possibilities were included in the analysis of \cite{Lin_Kratter2016} via their parameter $\theta$, although they did not interpret this parameter in terms of whether the external heating was held constant per unit mass or per unit area.

\subsection{Thermal Balance}

In the presence of self-gravity, an irradiated disc settles into a `thermally saturated' state, where radiative cooling and external irradiation is balanced by heating generated through gravitational instability. This heating represents the conversion of mechanical energy into heat as a result of shock dissipation associated with spiral arms in the disc. \cite{Lodato2004} demonstrated using global calculations that the relationship between such heating and the rate of associated angular momentum transfer is analogous to the action of an effective viscosity and so can be cast in terms of an equivalent value of the Shakura-Sunyaev $\alpha-$viscosity \citep{Shakura_Sunyaev1973}. Within this framework, the equation of thermal balance can be expressed in the form \citep{Rice2011}:

\begin{equation} \label{eq:alpha_beta}
\alpha = \frac{4}{9\gamma(\gamma-1)\beta}\left(1-\frac{Q_\text{{irr}}^2}{Q_\text{{sat}}^2}\right)
\end{equation}
This equation assumes $\langle \Sigma c_{\text{s}}^2 \rangle = \langle \Sigma \rangle \langle c_{\text{s}} \rangle^2$ and applies to both irradiation models ($f_\theta = 0,1$). It is only semi-analytic since the $Q$ value of the saturated system, $Q_{\text{sat}}$, is not known a priori and is dependent on the imposed level of irradiation, parametrised by $Q_{\text{irr}}$. The classical understanding is that for low levels of irradiation ($Q_{\text{irr}}\lesssim1$), a gravitoturbulent state with finite $\alpha$ can be maintained. However, once the level of irradiation implies a temperature such that $Q_{\text{irr}} > Q_{\text{sat}}$, $\alpha$ reduces to zero and the instability is inactive.

\cite{Rice2011}, using  $f_\theta=1$ in equation \eqref{eq:Energy_eq}, find an approximately constant value of $Q_{\text{sat}}$ with varying $Q_{\text{irr}}$ and so fix this when estimating $\alpha$ via equation \eqref{eq:alpha_beta}. They make the above assumption of stability under high irradiation and do not simulate $Q_{\text{irr}} > Q_{\text{sat}}$. The simulations in \cite{Lohnert2020} use $f_\theta=0$ and find that $Q_{\text{sat}}$ varies with $Q_{\text{irr}}$, such that it is always greater than $Q_{\text{irr}}$ and $\alpha$ remains finite. This includes $Q_{\text{irr}} > Q_{\text{sat,0}}$, where $Q_{\text{sat,0}}$ is that of the unirradiated ($Q_{\text{irr}}=0$) state. However, \cite{Lohnert2020} were unable to model a gravitoturbulent state with $Q_{\text{irr}}\gtrsim2 Q_{\text{sat,0}}$ on account of heating associated with the finite numerical viscosity in their code.

\subsection{Linear Stability Analysis}
\label{sec:linear_analysis}

We now perturb equations (\ref{eq:continuity_eqn}-\ref{eq:Poisson_eqn}) about a uniform equilibrium with constant density and pressure and corresponding Toomre $Q$: $Q_0=\frac{c_{\text{s,0}}\Omega_0}{\pi G \Sigma_0}$. We assume linear perturbations of the form $f'(\mathbf{x},t)=\tilde{f}e^{(\sigma t+ikx)}$, where $\sigma$ is the growth rate and $k$ is the $x$-component of the wave vector. We use axisymmetric perturbations for simplicity here, though non-axisymmetric perturbations are able to grow at larger $Q$ \citep{Lau1978,Papaloizou1989,Papaloizou1991,Durisen2007,Mamatsashvili2007}, which will affect the comparison between theory and simulation results. Here, we use $Q_{0} = Q_{\text{irr}}$ as the background state. Using this in the above equations leads to the dispersion relation:

\begin{equation} \label{eq:Lonhert_dispersion}
\sigma^2=-\Omega_0^2+\frac{2 c_{\mathrm{s}, 0} \Omega_0}{Q_{\text{irr}}} k-\frac{c_{\mathrm{s}, 0}^2(f_\theta +\gamma \tau_{\mathrm{c}} \sigma)}{\gamma(1+\tau_{\mathrm{c}} \sigma)} k^2
\end{equation}

This equation is equivalent to that found by \cite{Lin_Kratter2016}. When $f_\theta =1$, corresponding to constant heating per unit mass, unstable modes ($\sigma^2>0$) are possible for $\frac{\gamma}{Q_{\text{irr}}}\left(1-\sqrt{1-\frac{Q_{\text{irr}}^2}{\gamma}}\right)<\frac{kc_{s,0}}{\Omega_0}<\frac{\gamma}{Q_{\text{irr}}}\left(1+\sqrt{1-\frac{Q_{\text{irr}}^2}{\gamma}}\right)$. There are no unstable modes when $Q_{\text{irr}}>\sqrt{\gamma}$\footnote{This criterion differs from the standard Toomre criterion due to our use of the adiabatic sound speed when defining $Q$. As the growth rate approaches zero, the cooling timescale scale becomes much shorter than the growth timescale and perturbations behave isothermally. Defining $Q$ by the isothermal sound speed accounts for the factor of $\sqrt{\gamma}$ and recovers the $Q<1$ criterion.}, so in this regime, a large enough irradiation temperature will completely suppress GI. This is similar to the behaviour of the dispersion relation of a barotropic fluid without cooling, which has instability below a critical $Q$ and a larger range of unstable wavenumbers as $Q$ is decreased below this value.

Conversely, $f_\theta=0$ corresponds to constant heating per unit area and, as shown in \cite{Lohnert2020}, unstable modes are possible for all values of $Q_{\text{irr}}$ at high wavenumbers (small spatial scales): $\frac{kc_{s,0}}{\Omega_0}>\frac{Q_{\text{irr}}}{2}$.

\begin{figure}
    \centering
    \includegraphics[width=\columnwidth]{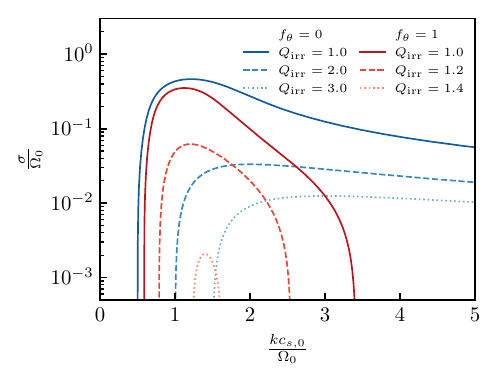}
    \caption{Analytic dispersion relations from equation \eqref{eq:Lonhert_dispersion} with cooling timescale $\beta=\tau_{\text{c}}\Omega_0=10$ and different levels of irradiation, $Q_{\text{irr}}$. Both the heating per unit area ($f_\theta=0$) and per unit mass ($f_\theta=1$) models are plotted.}
    \label{fig:Dispersion_relation}
\end{figure}

We plot  the dispersion relation in Fig. \ref{fig:Dispersion_relation} for different levels of irradiation using $\gamma=2$.

The most unstable mode can be found by solving $\partial \sigma/\partial k = 0$. For the high irradiation regime of interest, the growth rate becomes small. Neglecting higher orders of $\frac{\sigma_{\text{m}}}{\Omega_0}$,

\begin{equation} \label{eq:Reduced_Most_unstable_g}
\frac{\sigma_{\text{m}}}{\Omega_0} =\frac{\gamma - Q_{\text{irr}}^2f_\theta}{(Q_{\text{irr}}^2-1)\gamma\beta}
\end{equation}
Similarly, the expression for the most unstable wavenumber which is valid for $Q_{\text{irr}}\gtrsim1$ is given by

\begin{equation} \label{eq:Most_unstable_k}
\frac{k_{\text{m}}c_{s,0}}{\Omega_0} = Q_{\text{irr}}
\end{equation}
This expression suggests that with higher levels of irradiation, instability is expected to occur on shorter length scales. 

For $Q_{\text{irr}}<1$, the growth rate is too large to neglect higher orders. In this regime, $k_{\text{m}}c_{s,0}/\Omega_0 = 1/Q_{\text{irr}}$, which is equivalent to the standard Toomre criterion.

Following the analysis of \cite{Lohnert2020}, the growth rate of the most unstable mode can be used to derive analytic predictions for the saturated value of $Q$ and corresponding $\alpha$ of the turbulent state using a mixing length approach \citep{Shakura2018}. An estimate for the turbulent viscosity is found using characteristic mixing length and time scales given by a typical wavelength and growth rate respectively. Here, we keep $Q_0=Q_{\text{irr}}$, whereas \cite{Lohnert2020} use $Q_0=Q_{\text{sat}}$ to evaluate the growth rate. The more physically reasonable choice is somewhat unclear. We use $Q_0=Q_{\text{irr}}$ in the linear analysis as the background state is determined by the level of irradiation. However, the mixing length model is used for the saturated state, where it may be more appropriate to evaluate the growth rate using $Q_{\text{sat}}$ rather than $Q_{\text{irr}}$. We explored both options and found that $Q_0=Q_{\text{irr}}$ gave a closer fit to the data, although the difference in the resulting analytic expressions was minimal. Using this with the maximum growth rate and retaining higher orders results in the following equation, relating $\alpha$ to $Q_{\text{irr}}$ and $\beta$ (see Appendix \ref{sec:Mixing_length_appendix} for details):

\begin{equation} \label{eq:alpha_Qirr}
\begin{split}
& \gamma \beta Q_{\text{irr}}^2 \left(\frac{Q_{\text{sat}}^2}{Q_{\text{irr}}^2}\frac{\left( \frac{3}{\pi} \right)^2\alpha }{\Omega_0}\right)^3 +f_\theta Q_{\text{irr}}^2\left(\frac{Q_{\text{sat}}^2}{Q_{\text{irr}}^2}\frac{\left( \frac{3}{\pi} \right)^2\alpha }{\Omega_0}\right)^2 \\
& + \gamma \beta \left(\frac{Q_{\text{sat}}^2}{Q_{\text{irr}}^2}\frac{\left( \frac{3}{\pi} \right)^2\alpha }{\Omega_0}\right) \left(Q_{\text{irr}}^2-1\right) + (f_\theta Q_{\text{irr}}^2 - \gamma) =0
\end{split}
\end{equation}

In addition to the linear theory, the assumption of thermal equilibrium (equation \eqref{eq:alpha_beta}) allows us to find equations for $\alpha$ and $Q_{\text{sat}}$ as functions of $Q_{\text{irr}}$ and $\beta$ only. For $f_\theta=0$, this solution is physical down to $Q_{\text{irr}}=1.8$ with $Q_{\text{sat}}=3.2$. For low values of $Q_{\text{irr}}$, the growth rate becomes large and $Q_{\text{sat}}$ reduces to zero as the linear theory breaks down. Furthermore, in the low irradiation regime, the most unstable wavenumber is inversely proportional to $Q_{\text{irr}}$ as mentioned above, resulting in an unphysically short mixing length estimate as $Q_{\text{irr}}$ becomes small. It is expected theoretically and seen in numerical calculations that a disc in the absence of irradiation will sustain a quasi-steady state and approximately constant $Q_{\text{sat,0}}>1$ \citep{Paczynski1978}. With this in mind, we take $Q_{\text{irr}}=1.8$, where the linear theory diverges from the physical expectation, to be the limit of where it is a representative approximation. Below this, we assume a constant $Q_{\text{sat,0}}=3.2$.

To summarise the complete analytic model to be compared to the simulation results in Section \ref{sec:results}, we use equations \eqref{eq:alpha_beta} and \eqref{eq:alpha_Qirr} to predict $\alpha$ and $Q_{\text{sat}}$ for $f_\theta=0$ and $Q_{\text{irr}}>1.8$. For $Q_{\text{irr}}<1.8$ we take $Q_{\text{sat}}=Q_{\text{sat,0}}$ as constant and use only equation \eqref{eq:alpha_beta}.

For $f_\theta=1$, there are no solutions in the linear theory for $Q_{\text{irr}}>\sqrt{\gamma}$, which is lower than the $Q_{\text{irr}}=1.8$ threshold so the final model assumes $Q_{\text{sat}}=Q_{\text{sat,0}}$ until $Q_{\text{irr}}>Q_{\text{sat,0}}$. Above this, $Q_{\text{sat}}=Q_{\text{irr}}$ as the temperature is set completely by the irradiation.

\section{Methods}
\label{sec:Methods} 

\subsection{The Code}

The simulations in this work were performed with \textit{Athena} \citep{Stone2008}, a grid-based, second-order Godunov code. We solve the equations of hydrodynamics using the shearing sheet implementation \citep{Stone2010} with periodic boundary conditions in the $x$ and $y$ directions, accounting for the shearing motion at the radial boundaries. Self-gravity is included by solving for the potential via Fast Fourier Transforms.

The simulation domain is a grid of $N\times N$ cells with $N=1024$. Lengths in code units can be scaled by $H=\frac{\pi G \Sigma_0}{\Omega_0^2}$ and setting $\Sigma_0=\Omega_0=1$ and $G=1/\pi$, such that $H=1$. We use a box size of $L_x = L_y = 64H$ and choose an adiabatic index of $\gamma=2$ for comparison with the literature e.g. \cite{Gammie2001}. The pressure scale height of the disc is $H_{\text{p}}=QH$. The simulations were initialised with uniform density and pressure and $Q=1.1$.

Perturbations were introduced to the velocity field in the form of a Gaussian random field of amplitude $\langle\delta v^2\rangle/c_{\text{s}}^2=(1.0/1.1)^2$, following the method of \cite{Johnson2003}. This amplitude means that the simulations begin with non-linear perturbations. This was done to reduce computational time as for large values of $Q_{\text{irr}}$, the growth rate becomes small (equation \eqref{eq:Reduced_Most_unstable_g}), requiring many orbital timescales to reach the saturated state. With order unity perturbations, all simulations saturate within $50\Omega_0^{-1}$. To validate this approach, additional simulations were conducted with amplitudes of $\langle\delta v^2\rangle/c_{\text{s}}^2=(10^{-4}/1.1)^2$, across a range of $Q_{\text{irr}}$. The growth of perturbations in the linear phase was consistent with the analytic growth rate. Once the simulations reached a saturated state, there was no difference in their behaviour or measured properties compared to those initialised with order unity perturbations.

The $\alpha$-parameter can be measured directly from simulations via the gravitational and Reynolds stresses:

\begin{equation} \label{eq:alpha_grav_rey}
\alpha = \frac{2}{3\left\langle\Sigma c_s^2\right\rangle}\langle S_{xy} \rangle = \frac{2}{3\left\langle\Sigma c_s^2\right\rangle}\left(\langle G_{x y} \rangle + \langle H_{x y} \rangle \right)
\end{equation}
The Reynolds stress is calculated by 
\begin{equation} \label{eq:Rey_stress}
\langle H_{x y} \rangle=\langle \Sigma u'_x u'_y \rangle  \;, 
\end{equation}
where $u'_x$ and $u'_y$ are the velocities relative to the background orbital motion. The gravitational stress component is given by equation \eqref{eq:grav_stress}, which can be calculated analytically in the Fourier domain \citep{Gammie2001}.
\begin{equation} \label{eq:grav_stress}
G_{x y}=\int_{-\infty}^{\infty} d z \frac{g_x g_y}{4 \pi G}
\end{equation}
\begin{equation} \label{eq:Gk}
\left\langle G_{x y}\right\rangle=\sum_{\boldsymbol{k}} \frac{\pi G k_x k_y\left|\Sigma_k\right|^2}{|\boldsymbol{k}|^3}
\end{equation}

Initial results were found to behave as expected without the addition of artificial viscosity. Given the low values of $\alpha$ expected in the highly irradiated regime, added viscosity would dominate and the contribution from GI would be undetectable. For these reasons, no artificial viscosity is included in the simulations.

\subsection{Fragmentation Criterion}
\label{sec:frag_crit}
After a saturated state is achieved, fragmentation can occur by the runaway collapse of overdense regions. This happens when $\beta$ is less than a critical cooling timescale, $\beta_{\text{crit}}$, below which overdensities are unable to be stabilised by heating. The precise location of this boundary has faced issues due to nonconvergence for a variety of reasons associated with the dimensionality and viscosity implementation in the codes \citep{Lodato2011,Paardekooper2011,Meru2012}, but is typically quoted as $\beta_{\text{crit}}\approx3-10$ \citep{Gammie2001,Rice2005}.

There is no criterion for identifying fragmentation that has been uniformly applied in the literature, although a common choice requires an overdensity above $100\Sigma_0$ to survive for a few orbital timescales \citep{Meru2011,Rice2011,Paardekooper2012}. From initial results, classifying simulations as fragmented was found to depend sensitively on the exact threshold for the maximum density and survival time of a clump. Using the condition of $\Sigma_{\text{max}}>100\Sigma_0$ for $t>3\Omega_0^{-1}$ indicated fragmentation in cases which were later found to contain only transient overdensities. To distinguish these simulations from fragmenting ones, we use a further criterion that fragments must be gravitationally bound. 

\subsection{Time-dependent cooling}
\label{sec:time-dependent_cooling}
Simulations are initiated with $\beta=10$ and allowed to settle into a saturated state. When simulating shorter cooling times, the value of $\beta$ is then linearly reduced over $100\Omega_0^{-1}$ to its target value. If low values of $\beta$  are introduced at $t=0$, the disc may spuriously fragment in this initial phase. As pointed out by \cite{Clarke2007}, it is also not realistic to instantaneously form a disc with $\beta < \beta_{\text{crit}}$, but the disc may evolve to have a lower cooling time due to infall of material or redistribution via gravitational torques. They study the effect of slowly reducing $\beta$ over many cooling times and find this lowers the value of $\beta_{\text{crit}}$ by a factor of ~2 compared to a rapid decrease\footnote{This was tested in the non-irradiated case, where we observed that a simulation with $\beta=3$ fragmented upon an instantaneous reduction of the cooling time, but did not fragment when gradually reduced}. The dependence on thermal history should be accounted for when comparing values of $\beta_{\text{crit}}$. Averaged quantities are calculated from after the final $\beta$ has been reached.

\subsection{Resolution Effects}

The standard set-up for simulations used a box size of $L_x=L_y=64H$ and $N=1024$. Both box size and resolution can affect the outcome of simulations and absolute quantities \citep{Young2015,Riols2017,Booth2019}. Using the non-irradiated case $\left(Q_{\text{irr}}=0\right)$ and $\beta=10$, simulations with $N=256,512,1024,2048$ were conducted. There was minimal effect on quantities of interest, such as average $Q$ and $\alpha$, which were found to have converged by $N=1024$. Table \ref{tab:res_box_size} provides a summary of results from tests of different simulation configurations.

\begin{table}
    \caption{Averaged properties over simulation domain and time for simulations with $f_{\theta}=0$ and $\beta=10$. }
    \centering
    \begin{tabular}{llccc} 
    \hline
          $Q_\text{irr}$&Box size &Resolution &  $Q_\text{sat}$& $\langle\alpha\rangle$\\
 & $(L_x\times L_y)$& $(N_x\times N_y)$& &\\
 \hline
 0 & $8H\times 8H$& $1024\times 1024$ & 1.67&0.0241\\
 0 & $16H\times 16H$& $1024\times 1024$ & 2.10&0.0236\\
 0 & $32H\times 32H$& $1024\times 1024$ & 2.68&0.0210\\
 0 & $64H\times 64H$& $256 \times 256$ & 3.16&0.0221\\
 0 & $64H\times 64H$& $512 \times 512$ & 3.00&0.0218\\
 0 & $64H\times 64H$& $1024\times 1024$ & 2.96&0.0216\\
 0 & $64H\times 64H$& $2048\times 2048$ & 2.89&0.0215\\
 0 & $128H\times 128H$& $1024\times 1024$ & 2.97&0.0217\\
 10 & $64H\times 64H$& $512\times 512$ & 10.26&0.00102\\
 10 & $64H\times 64H$& $1024\times 1024$ & 10.28&0.00109\\
 10 & $128H\times 128H$& $1024\times 1024$ & 10.36&0.00135\\ 
 \hline
    \end{tabular}
    
    \label{tab:res_box_size}
\end{table}

With no irradiation, $Q$ increases quickly from its initial value and settles to $Q_{\text{sat,0}}\approx3.0$. Including irradiation leads to $Q$ saturating at a higher value as $Q_\text{irr}$ is increased, as seen in Fig. \ref{fig:Qvst}. Previous results using ZEUS \citep{Mamatsashvili2009} and Athena \citep{Shi2016} have also found these codes to have higher values of $Q_{\text{sat,0}}$ than others, which more commonly saturate at $Q_{\text{sat}}\approx2.0$. Many numerical factors can increase $Q_\text{{sat}}$, such as using a large box size \citep{Riols2017,Booth2019,zier2023}. The $Q_{\text{irr}}=0$ case was tested with $L_x=L_y=8,16,32,64,128H$.  There was little difference to the average $\alpha$ values, but $Q_{\text{sat}}$ was found to increase with box size (see Table \ref{tab:res_box_size}). For small boxes, $Q_{\text{sat}}\approx1.7$, which is closer to that measured in similar studies \citep{Rice2011}. Convergence was reached for $L\geq64H$. However, using a box of this size limits the applicability of the local model to real astrophysical discs.

All simulations were initiated with $Q=1.1$, implying a scale height $H_{\text{p}}=1.1$. However, as they evolve, $H_{\text{p}}$ increases with $Q$, changing the size of the box in units of $H_{\text{p}}$ and the resolution in units of cells per pressure scale height. In the non-irradiated case and in previous studies, $Q_{\text{sat}}$ remains low enough such that the difference is small. In Section \ref{sec:Nonlinear_State} we demonstrate that in the highly irradiated regime, $Q_{\text{sat}}$ is large enough to increase the scale height by a factor of 10 relative to initial conditions. Although this only increases the number of cells per scale height, it means that simulation with different inputs of $Q_{\text{irr}}$ yield different $H_{\text{p}}$. To assess whether this influences the conclusions, $Q_{\text{irr}}=10$ simulations were performed with larger box sizes and/or lower resolutions in order to match the box size and number of cells per $H_{\text{p}}$ to the $Q_{\text{irr}}=0$ case. There was minimal effect on the simulation properties and behaviour (Table \ref{tab:res_box_size}), supporting the use of the standard set-up with $N=1024$ and $L=64H$ for subsequent simulations. Considering also a theoretical condition, the box length should accommodate the length scale corresponding to the minimum unstable wavenumber from the linear dispersion relation (equation \ref{eq:Lonhert_dispersion}) i.e. $L>\frac{2\pi}{k_{\text{min}}}=\frac{4\pi}{Q_{\text{irr}}}H$. This is satisfied for all but the smallest values of $Q_{\text{irr}}$, where $k_{\text{min}}$ becomes very small. However, in practice $Q$ always saturates to values at least as high as $Q_{\text{sat,0}}\sim3$, such that very small wavenumbers are never realised.

\section{Results and Discussion}
\label{sec:results}

\subsection{Nonlinear State}
\label{sec:Nonlinear_State}
\begin{figure*}
\centering    
\includegraphics[width=\textwidth]{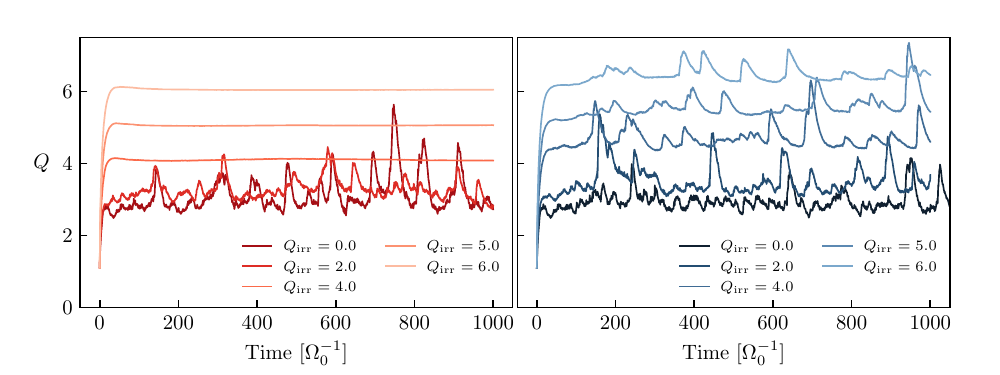}
\caption[Qvst]{Temporal evolution of the domain-averaged $Q$ for a selection of simulations with different levels of irradiation, $Q_{\text{irr}}$. \textbf{Left}: $f_\theta=1$. \textbf{Right}: $f_\theta=0$.}
\label{fig:Qvst}
\end{figure*}

Simulations were run across a range of input parameters, including cooling times $\beta=10,20$, and irradiation level, $Q_{\text{irr}}=$ \numrange[range-phrase = --]{0}{10}, for both the case of heating per unit area and per unit mass. In all simulations, $Q$ increased sharply within a few $\Omega_0^{-1}$ from an initial condition of $Q=1.1$ and either saturated to a gravitoturbulent state or resulted in laminar flow. Example time series of $Q$ are shown in Fig. (\ref{fig:Qvst}) for $\beta=10$. The cooling time was chosen such that fragmentation does not occur and a thermally saturated state can be achieved. The fragmentation boundary is investigated using lower $\beta$ values in Section \ref{sec:Frag_boundary}. 

\begin{figure}
\centering    
\includegraphics[width=\columnwidth]{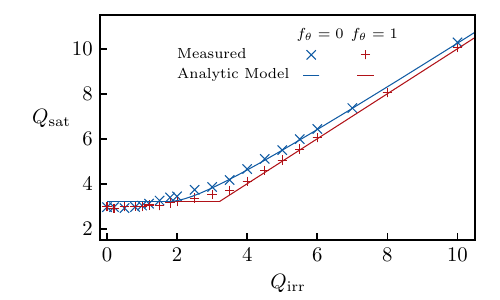}
\caption[QsatvsQirr]{Saturated values of $Q$ measured from simulations with $\beta=10$. $Q_{\text{sat}}$ is calculated by averaging over time and the simulation domain during the saturated state. The solid lines indicate the analytic model outlined in Section \ref{sec:linear_analysis}.}
\label{fig:Qsat}
\end{figure}

 With no irradiation, $f_\theta=0$ and $f_\theta=1$ are equivalent and $Q$ settles to $Q_{\text{sat,0}}\approx3.0$. This agrees well with the value from the analytic model in Section \ref{sec:linear_analysis} ($Q_{\text{sat,0}}=3.2$), despite being larger than what is commonly measured in simulations. Increasing $Q_{\text{irr}}$ raised the value of $Q$ reached by the simulations. For $f_\theta=1$, $Q$ remained constant over time for higher levels of iradiation, whereas for $f_\theta=0$, $Q$ continued to fluctuate over time, as seen in fig. \ref{fig:Qvst}.

Fig. \ref{fig:Qsat} shows how the saturated value of $Q$ varies with the imposed level of irradiation. For $f_\theta=1$, $Q_{\text{sat}}$ remains approximately constant with increasing $Q_{\text{irr}}$ until $Q_{\text{irr}}>Q_{\text{sat,0}}$. The instability becomes inactive and $Q$ terminates at $\approx Q_{\text{irr}}$. The resulting laminar state is demonstrated by the flat $Q$ vs time profiles, low $\alpha$ values and the spatial structure of density.

For $f_\theta=0$, $Q_{\text{sat}}$ remains approximately constant for $Q_{\text{irr}}<1.8$. In this regime, we expect the measured value of $Q_{\text{sat}}$ to be determined by non-linear effects in the simulations and to be largely independent of the low $Q_{\text{irr}}$ value. For $Q_{\text{irr}}>1.8$, irradiation becomes a significant contribution to the heating balance. The saturated state responds to the level irradiation, such that $Q_{\text{sat}}$ increases, remaining greater than $Q_{\text{irr}}$ and the instability is still active. This behaviour is seen in the simulation results, which agree well with the model.

\begin{figure}
\centering    
\includegraphics[width=\columnwidth]{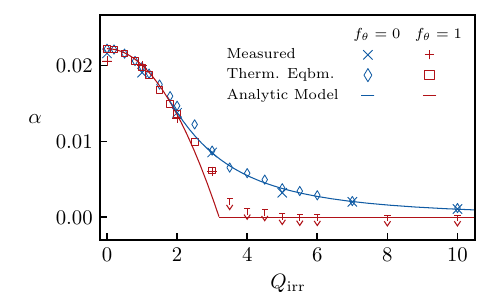}
\caption[$\alpha$ vs $Q_{\text{irr}}$ comparing theoretical and measured values]{Behaviour of $\alpha$ with increasing irradiation at $\beta=10$. Direct measurements from simulations using equation \eqref{eq:alpha_grav_rey} are compared to calculations of $\alpha$ using the assumption of thermal equilibrium (equation \eqref{eq:alpha_beta}) with $Q_{\text{sat}}$ measured directly from the simulations. Theoretical lines are plotted using the linear analysis of Section \ref{sec:linear_analysis}. Anomalous $\alpha$ measurements are plotted as upper limits.}
\label{fig:alpha}
\end{figure}

The $\alpha$-parameter is a measure of the efficiency of angular momentum transport due to turbulence and indicates the activity of GI. It was measured directly from gravitational and Reynolds stresses in the simulations via equation \eqref{eq:alpha_grav_rey} and compared to theoretical predictions outlined in Section \ref{sec:linear_analysis}. Fig. \ref{fig:alpha} plots the variation of $\alpha$ with $Q_{\text{irr}}$. The behaviour is consistent with Fig. \ref{fig:Qsat} with the cases $f_\theta=0$ and $f_\theta=1$ diverging around $Q_{\text{irr}}=1.8$. For $f_\theta=1$, $\alpha$ reduces to zero as irradiation dominates the saturated state. For $f_\theta=0$, $\alpha$ remains finite for high levels of irradiation.

Though the theory predicts no unstable solutions for $f_\theta=1$ and $Q_{\text{irr}}>Q_{\text{sat,0}}$, residual $\alpha$ values were measured for simulations at higher $Q_{\text{irr}}$. This is due to the implementation of self-gravity in the code not fully conserving momentum. The disc acquires a non-zero average velocity in the form of a small eccentricity, oscillating between the radial and azimuthal directions. The resulting Reynolds stress is due to this bulk motion rather than turbulent motion associated with GI. Other properties of these simulations indicate that GI is not active, such as the maximum density remaining below $\sim2\Sigma_0$. For this reason, we mark these values as upper limits in Fig. \ref{fig:alpha}.

The agreement between the directly measured $\alpha$ and calculating via equation \eqref{eq:alpha_beta} supports the assumption that thermal equilibrium has been reached and shows the simulation is conserving energy well. There is also excellent agreement between the simulations and the analytic model derived from the linear theory.

\begin{figure*}
\centering    
\includegraphics[width=\textwidth]{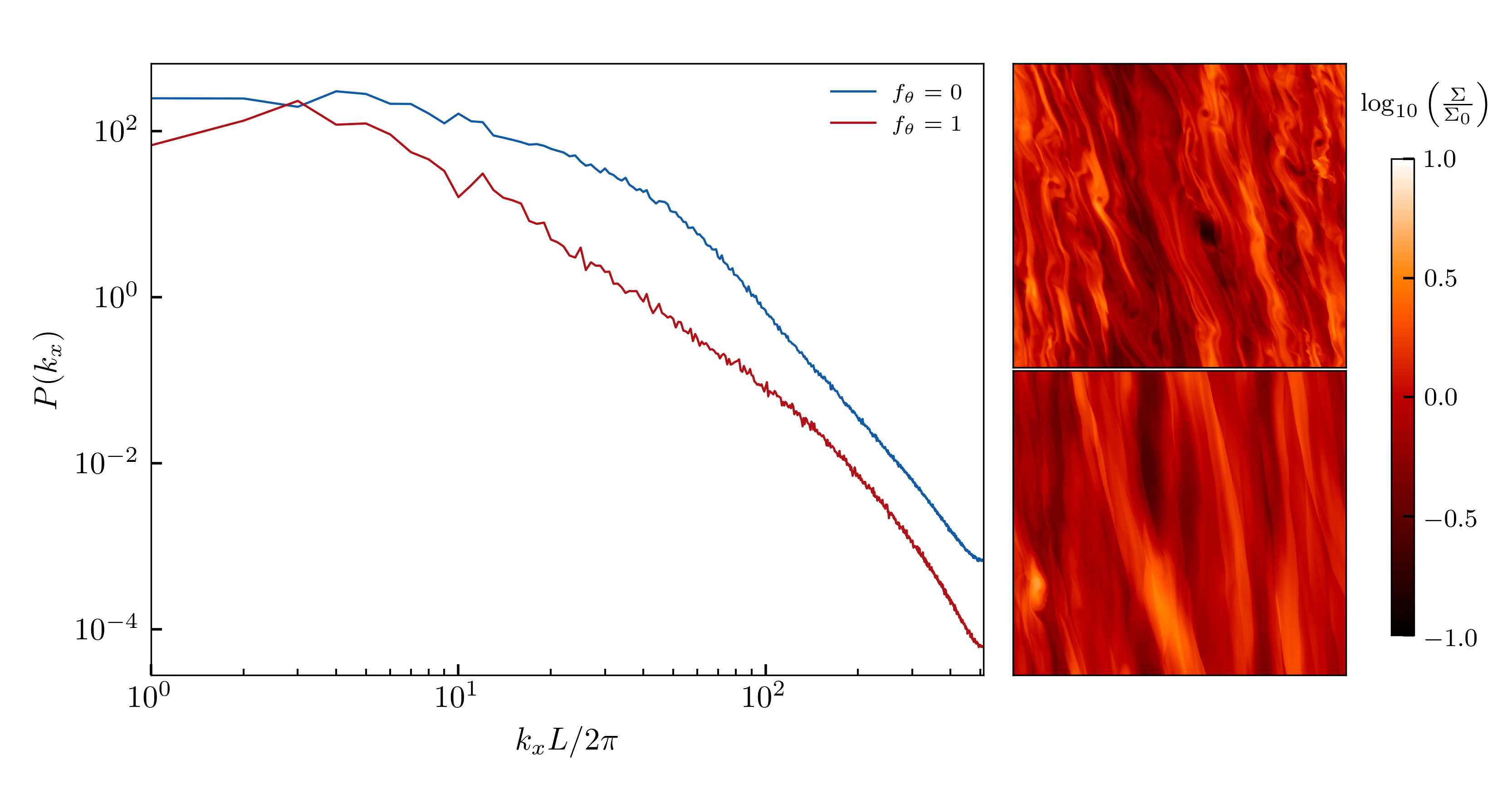}
\caption[Power_spectra]{\textbf{Left}: Power of density perturbations as a function of $k_x$ for simulations with $Q_{\text{irr}}=3.0$ and $\beta=4$. Both simulations are in a gravitoturbulent state with $\alpha=0.025$ for $f_\theta=0$ and $\alpha=0.0089$ for $f_\theta=1$. \textbf{Right:} Surface density snapshots of the same simulations for $f_\theta=0$ (top) and $f_\theta=1$ (bottom).}
\label{fig:power_spectra}
\end{figure*}

From the linear analysis, Fig. \ref{fig:Dispersion_relation} suggests that irradiation will limit the range of unstable wavenumbers for $f_\theta=1$ compared to $f_\theta=0$. Fig. \ref{fig:power_spectra} shows example power spectra for each irradiation model. Following the method of \cite{Lohnert2020}, Fourier amplitudes of density perturbations were computed at a given timestamp and averaged over the y-component of the wavenumber to obtain the power as a function of the x-component. These were then averaged over multiple timestamps during the saturated state: $P(k_x) = \left\langle \langle|\hat{\Sigma}(k_x,k_y)|^2\rangle_{k_y} \right\rangle_t$. Fig. \ref{fig:power_spectra} also includes example snapshots of the spatial structure of density. The $f_\theta =0$ simulation exhibits more small-scale structure, which is evident in the power spectra. The shape of the power spectra differ between the models, with $f_\theta=0$ remaining a shallower function of $k_x$ until $\frac{k_x L}{2\pi}\approx 30$. The greatest difference in power occurs around this wavenumber, corresponding to wavelengths of order $H$ and the wavenumber of the most unstable mode predicted by linear theory. At large wavenumbers, pressure is the dominant restoring force. When irradiation is supplied per unit area of the disc, these sub-Jeans length scales are less effectively stabilised, resulting in the retention of power at small scales.

\subsection{Fragmentation}
\label{sec:Frag_boundary}

The process of fragmentation can be considered in two stages. Firstly, the disc must be gravitationally unstable, for which we have shown is possible for $f_\theta=0$ heating even at high levels of irradiation. Secondly, the disc must be able to cool sufficiently quickly that overdensities are not able to be stabilised by heating associated with PdV work and shock heating and instead continue to grow. Otherwise, the disc saturates to the gravitoturbulent state. Due to the dependence of fragmentation on cooling time, the criterion for fragmentation is often considered as a critical value for the $\beta$-parameter, such that fragmentation occurs for $\beta < \beta_{\text{crit}}$.

There has been much uncertainty over the location and nature of this boundary as values of $\beta_{\text{crit}}$ are not consistent across the literature or are unable to converge due to resolution dependence \citep{Meru2011,Lodato2011,Meru2012}. In the absence of irradiation, we find fragmentation below $\beta_{\text{crit}} \approx 2$, which is similar to the value of $\beta_{\text{crit}} =3$ found by \cite{Gammie2001}, but lower than $\beta_{\text{crit}} = 8$ found by \cite{Rice2011}. The difference may be due to exact numerical implementations used and the equation of state \citep{Rice2005}. \cite{Rice2011}'s value of $\gamma=1.6$ compared to $\gamma=2$ here would result in a factor of two difference in equation \eqref{eq:alpha_beta}. $\beta_{\text{crit}}$ will also be lowered by a factor of $\approx 2$ by the decision to slowly reduce the cooling time \citep{Clarke2007}. Similarly, \cite{Young2015} investigate the dependence of $\beta_{\text{crit}}$ on softening and resolution. They report $\beta_{\text{crit}} \approx 7$ for unsoftened simulations using $\gamma = 5/3$ and at a similar resolution to this work.

Given the lack of softening, we do not expect the absolute values for the fragmentation boundary to numerically converge. We instead focus on running simulations with consistent properties, such as resolution and box size in order to analyse the functional behaviour of $\beta_{\text{crit}}$ with $Q_{\text{irr}}$. Furthermore, the radial dependence of $\beta$ in the outer disc ($\beta \propto r^{-9/2}$ \citep{Clarke2009, cossins2010}) is steep enough that small differences in $\beta_{\text{crit}}$ do not correspond to significant changes in radial location.

\begin{figure*}
    \centering
    \includegraphics[width=\textwidth]{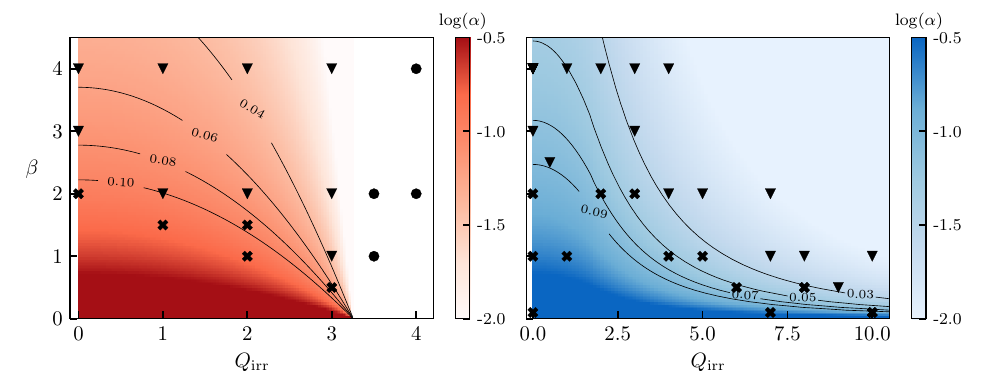}
    \caption{Outcome of simulations with different cooling times and levels of irradiation. Crosses and triangles represent fragmented and unfragmented simulations respectively. Circles indicate simulations with no gravitational instability. \textbf{Left}: $f_\theta=1$. \textbf{Right}: $f_\theta=0$. The colorbar shows the theoretical value of $\alpha$ using the analytic model in Section \ref{sec:linear_analysis}.}
    \label{fig:Frag_Boundary}
\end{figure*}

Simulations were conducted following the previous methodology, but with a range of $\beta$ values. Fig. \ref{fig:Frag_Boundary} shows the three possible outcomes in the $(\beta,Q_{\text{irr}})$ parameter space:  

\hspace{5mm}

i) Gravitationally unstable, but not fragmenting. There was clear turbulent activity, but overdensities did not grow and persist.

\hspace{2mm}

ii) Gravitationally unstable and fragmented. Simulations reached extreme maximum densities and formed bound fragments.

\hspace{2mm}

iii) Gravitationally stable. No turbulent activity observed. The maximum density does not increase beyond a few times the background density.

\hspace{5mm}

The fragmentation boundary is usually determined in simulations without irradiation, for which $\beta\sim3$ is found, corresponding to $\alpha\sim0.07$ \citep{Gammie2001,Rice2005,Paardekooper2012}. The simulations here show $\beta_{\text{crit}}$ to be a decreasing function of $Q_{\text{irr}}$, suggesting it may be more suitable to express the fragmentation boundary using a critical value of $\alpha$, $\alpha_{\text{crit}}$, which is independent of $Q_{\text{irr}}$. $\alpha_{\text{crit}}$ relates to the amplitude of density perturbations \citep{Cossins2009} and regulates the maximum stress the disc can withstand before fragmenting. Using the analytic model from Section \ref{sec:linear_analysis}, contours of constant $\alpha$ are plotted on Fig. \ref{fig:Frag_Boundary}.

\subsection*{Case $f_\theta=1$}

For the case of heating per unit mass, fragmentation only occurred for low levels of irradiation and with short cooling times. The value of $\beta$ required for collapse decreased as the irradiation was increased. For $Q_{\text{irr}} > Q_{\text{sat,0}}$, there was no evidence of instability due to self-gravity, so fragmentation could not occur even for very low cooling times. This behaviour follows the classical understanding that once the level of irradiation implies a corresponding value of $Q_{\text{irr}}$ that exceeds that of the natural saturated state of the system, GI will be quenched. The distinct change in behaviour as the instability is suppressed and $\alpha$ is reduced to zero occurs between $Q_{\text{irr}}=3.0-3.5$, in agreement with the analytic prediction of Section \ref{sec:linear_analysis}.

In the low irradiation regime with $Q_{\text{irr}} < Q_{\text{sat,0}}$, $\beta_{\text{crit}}$ decreases as $Q_{\text{irr}}$ increases. This boundary approximately follows the contours of constant $\alpha\sim0.08$, although there is not complete agreement. As was found in \cite{Rice2011}, the fragmentation boundary is such that a lower value of $\alpha$ is required for fragmentation as $Q_{\text{irr}}$ increases. 

\subsection*{Case $f_\theta=0$}

For the case of heating per unit area, GI remained active and fragmentation occurred at high levels of irradiation beyond the value of $Q_{\text{sat,0}}$. In this regime, the fragmentation boundary again approximately follows contours of $\alpha$, which, unlike $f_\theta=1$, only asymptotically approach zero. This allows a region of parameter space to be susceptible to fragmentation for high values of irradiation, up to the maximum $Q_{\text{irr}}=10$ tested. Here, the boundary occurs at a slightly lower $\alpha_{\text{crit}}\approx0.03$. This contour follows the fragmentation boundary until $Q_{\text{irr}}\approx3.5$, below which the threshold for $\alpha$ increases for decreasing values of $Q_{\text{irr}}$. At $Q_{\text{irr}}=0$, $\alpha_{\text{crit}}\approx0.09$.

As discussed in Section \ref{sec:frag_crit}, the observed boundary is sensitive to the chosen fragmentation criteria, particularly at low $Q_{\text{irr}}$. A more relaxed criterion results in higher $\beta_{\text{crit}}$ and lower $\alpha_{\text{crit}}$, which would be more consistent with a constant $\alpha$ threshold and with previously reported values of $\beta_{\text{crit}}$ for the non-irradiated case. Furthermore, in the high $Q_{\text{irr}}$ regime, the $\alpha$ contours converge significantly, leading to uncertainty in the value of $\alpha_{\text{crit}}$.

Determining an exact and consistent $\alpha_{\text{crit}}$, if one exists, will face similar numerical issues as those found for $\beta_{\text{crit}}$ and is not of significant consequence in locating the fragmentation zone of the disc owing to the steep radial dependence of $\beta$. The notable distinction here compared with $f_\theta=1$ is that fragmentation still occurs in highly irradiated regimes corresponding to $Q\sim10$. Both $Q_{\text{irr}}$ and $\beta$ decrease with radius, suggesting there is a radius in the disc beyond which $\alpha > \alpha_{\text{crit}}$ and fragmentation can occur.

\subsection*{Impact on Fragment Mass}
Fragmentation of a gravitationally unstable disc as a mechanism for planet formation faces problems in replicating expected planet masses. Population synthesis models indicate that planets formed via GI are overly massive when compared to the detection rates of direct imaging \citep{Forgan2018}. A simple estimate for the initial mass can be derived using the wavelength of the most unstable mode: $M_i \approx \Sigma \lambda^2$ \citep{Kratter2010}. In the standard theory, $\lambda = 2 \pi H Q$ with $Q\approx1$ required for fragmentation. Here, we have $\lambda = 2 \pi H/Q_{\text{irr}}$ from equation (\ref{eq:Most_unstable_k}). In the $f_\theta=0$ regime, irradiated discs with high $Q_{\text{irr}}$ are still able to fragment for sufficiently low $\beta$ and will do so on smaller length scales, leading to initial fragments of lower masses. 

Conversely, \cite{Cadman2020a} use the methods of \cite{Forgan2013} to calculate the Jeans mass in the spiral arms of irradiated, self-gravitating discs. They find that increased disc temperatures yield higher Jeans masses and so more massive fragments. Thorough analysis of fragment masses was not conducted in this study due to runtime considerations as the time-step required to accurately evolve the system reduces significantly as clumps form.

\section{Conclusions}
\label{sec:Conclusion} 
In this paper, we compared the analytic and numerical behaviour of a razor thin, self-gravitating protoplanetary disc using $\beta$-cooling and two models for heating due to irradiation. An important and under-explored regime where gravitational instability may occur in protoplanetary discs is the outer disc where cooling times are short, but where the disc is subject to significant heating by external irradiation. Our results highlight the sensitivity of gravitational instability to assumptions about how the incident heating rate responds to changes in surface density in the disc.

The two models include irradiation as either a constant heating rate per unit mass or per unit area of the disc, approximating heating of an optically thin and thick disc respectively. Irradiation alongside heating due to shocks in a gravitationally unstable disc can balance cooling, leading to a self-regulated state where the Toomre $Q$ parameter saturates to an approximately constant value.

When heating per unit mass, a gravitoturbulent state can be maintained only when irradiation does not imply a temperature, or corresponding $Q$, greater than that set by the saturated state of the system. In contrast, a constant heating rate per unit area means gravitoturbulence persists for higher levels of irradiation as the temperature of the saturated state is able to increase above that imposed by irradiation, in agreement with the analytic model. The Toomre $Q$ of discs in this regime is elevated, with gravitational instability still active for discs with $Q_{\text{sat}}\approx 10$, though with low values of the effective viscous stress, $\alpha$. This is in contrast to the conventional assumption that a gravitationally unstable disc must have $Q_{\text{sat}}\approx 1-2$. This result increases the parameter space of disc properties that may be susceptible to GI.

Gravitoturbulence relies on an active balance of heating and cooling processes. High irradiation levels will increase the value of $Q$ and decrease the viscous stress, whereas shorter cooling timescales will increase the viscous stress level but not affect the saturated $Q$ value (see Fig. \ref{fig:Qsat} and \ref{fig:alpha}). As regions of gas cool and collapse, they undergo more dissipative heating due to turbulent motions. When supplying irradiation per unit mass of the disc, overdense regions are preferentially heated and stabilised. This is not the case when heating per unit area and radiative cooling is still effective in balancing heating. Young protoplanetary discs are expected to be optically thick to incident irradiation, so it is more appropriate to prescribe irradiation as a constant heating rate per unit area. However, the heating and cooling functions used here are still idealised models for the disc thermodynamics.

This weak GI regime is important itself as it allows for angular momentum transport and the formation of spiral structures in discs. Although simulations here using heating per unit area remain unstable when highly irradiated, they have elevated temperatures and low values of viscous stress, which will reduce the amplitude and detectability of spiral waves \citep{Hall2018}. When including a dust component, the growth of solids can be enhanced as dust is concentrated in spiral arms \citep{Rice2005,Gibbons2012,Booth2016,Baehr2021,Baehr2022,Rowther2024dust,Longarini2023}, which can improve detectability \citep{Dipierro2015a,Cadman2020b}. 

To probe the conditions under which the gas component of an irradiated gravitoturbulent disc may fragment, low values of the cooling timescale were tested. The fragmentation boundary was found to be a function of irradiation, such that more irradiated discs require shorter cooling times in order to fragment. The boundary for fragmenting simulations approximately followed contours of constant $\alpha$ with simulations fragmenting for $\alpha \gtrsim 0.03-0.09$, consistent with previous estimates \citep{Gammie2001, Rice2005, Paardekooper2012}. In the case of heating per unit area, fragmentation can occur at values of $Q$ substantially greater than unity.

These results were obtained using a razor thin disc model with a simple cooling prescription. More realistic 3D models are required to test this further as the effect of self-gravity is generally reduced compared to the 2D approximation \citep{Mamatsashvili2010}. Including a more realistic cooling function or radiative transfer models will also be useful due to the dependence on thermodynamics of the results.

\section*{Acknowledgements}
CSL thanks the UK Science and Technology Facilities Council (STFC) Center for Doctoral Training (CDT) in Data Intensive Science at the University of Cambridge for a Ph.D. studentship and also thanks Kaitlin Kratter for useful discussions. RAB thanks the Royal Society for their support in the form of a University Research Fellowship. CJC has been supported by the Science and Technology Facilities Council (STFC) via the consolidated grant ST/W000997/1 and also by the European Union's Horizon 2020 research and innovation programme under the Marie Sklodowska-Curie grant agreement No. 823823 (RISE DUSTBUSTERS project). This work was performed using resources provided by the Cambridge Service for Data Driven Discovery (CSD3) operated by the University of Cambridge Research Computing Service (\url{www.csd3.cam.ac.uk}), provided by Dell EMC and Intel using Tier-2 funding from the Engineering and Physical Sciences Research Council (capital grant EP/T022159/1), and DiRAC funding from the Science and Technology Facilities Council (\url{www.dirac.ac.uk}).

\section*{Data Availability}

The code used for the simulations presented in this work (\textit{Athena}; \cite{Stone2008}) is available at https://princetonuniversity.github.io/Athena-Cversion/.



\bibliographystyle{mnras}
\bibliography{references}




\appendix

\section{Mixing length model}
\label{sec:Mixing_length_appendix}

The maximum growth rate from the linear analysis of Section \ref{sec:linear_analysis} can be used to estimate $\alpha$ and $Q_{\text{sat}}$ of the turbulent state using a mixing length approach \citep{Shakura2018}. 

Firstly, the turbulent viscosity can be estimated by characteristic time and length scales:
\begin{equation} 
\nu_t = \left\langle \frac{\delta_x^2}{\delta_t} \right\rangle
\end{equation}

We estimate the typical timescale as $\delta_t^{-1} = \sigma_{\text{m}}$, using the maximum growth rate derived from the dispersion relation (equation (\ref{eq:Lonhert_dispersion}) of the main text). The full expression for $\sigma_{\text{m}}$ is
\begin{equation}\label{eq:full_max_growth_rate}
\gamma \beta Q_{\text{irr}}^2 \left(\frac{\sigma_{\text{m}}}{\Omega_0}\right)^3 +f_\theta Q_{\text{irr}}^2\left(\frac{\sigma_{\text{m}}}{\Omega_0}\right)^2 + \gamma \beta \left(\frac{\sigma_{\text{m}}}{\Omega_0}\right) \left(Q_{\text{irr}}^2-1\right) + f_\theta Q_{\text{irr}}^2 - \gamma =0
\end{equation}

For the typical wavelength, \cite{Lohnert2020} find the power spectra of density perturbations to be insensitive to $Q_{\text{irr}}$. They average over it to find a typical wavenumber of $k\approx\frac{3}{2}$, leading to $\nu_t = \left( \frac{\pi}{3} \right)^2  \sigma_{\text{m}}$. 

The turbulent viscosity relates to the viscous stress, $\langle S_{xy} \rangle = \frac{3}{2}\langle\Sigma\rangle \nu_t$. Similarly, equation (\ref{eq:alpha_grav_rey}) relates the stress to the $\alpha$ parameter, leading to $\alpha = \frac{\langle \Sigma \rangle \nu_t}{\gamma(\gamma-1)\langle U\rangle}$. Applying equation (\ref{eq:alpha_beta}) allows us to eliminate $\langle U\rangle$. Here, \cite{Lohnert2020} note that $\nu_t$ is normalised with the background sound speed rather than $\langle c_{\text{s}}\rangle$ and as such, we use $\nu_t'=c_{\text{s,irr}}^2\nu_t$, resulting in $\alpha = \frac{\nu_t'}{1+\frac{\nu_t'}{\alpha_0}}$.

Substituting for the mixing length estimate of the turbulent viscosity gives $\sigma_{\text{m}}=\left( \frac{3}{\pi} \right)^2\frac{\alpha}{1-\frac{\alpha}{\alpha_0}}$, which can then be used with equation (\ref{eq:full_max_growth_rate}) to obtain equation (\ref{eq:alpha_Qirr}), relating $\alpha$ to $Q_{\text{irr}}$ and $\beta$.


\bsp	
\label{lastpage}
\end{document}